\documentclass[11pt]{article}
\usepackage{graphicx,rotating}

\newcommand{\BaBarYear}       {02}
\newcommand{\BaBarNumber}     {05}
\newcommand{\SLACPubNumber} {9170}

\newcommand{\BaBarType}     {CONF}  % Conference submission paper
\input mybabarsym

\def\absDe {\mbox{$| \Delta E |$}\xspace}
\def\absmes {\mbox{$| m_{ES}-m_B^{nominal} |$}\xspace}

\def\PM     {\ensuremath{\pm}\xspace}

\def\ksdk   {\ensuremath{\KS \to \pipi}\xspace}
\def\pizdk  {\ensuremath{\piz \to \gaga}\xspace}

\def\KKbar  {\ensuremath{\kaon \Kb}\xspace}
\def\golden {\ensuremath{\Bz \to \jpsi \Kz}\xspace}

\def\goldKP {\ensuremath{\Bp \to \jpsi \Kp}\xspace}
\def\etacK  {\ensuremath{\B \to \etac \kaon}\xspace}
\def\etacKN {\ensuremath{\Bz \to \etac \KS}\xspace}
\def\etacKn {\ensuremath{\Bz \to \etac \Kz}\xspace}

\def\etacKP {\ensuremath{\Bp \to \etac \Kp}\xspace}

\def\jpsiKstar  {\ensuremath{\B \to \jpsi \Kstar}\xspace}
\def\eGamEtac {\ensuremath{\jpsi \to \g \etac}\xspace}
\def\eFourK  {\ensuremath{\etac \to \KpKm \KpKm}\xspace}
\def\eKKKK   {\ensuremath{\etac \to ~4K}\xspace}

\def\eKKPi   {\ensuremath{\etac \to \KpKm \piz}\xspace}
\def\eKsKPi  {\ensuremath{\etac \to \KS \Kpm \pimp}\xspace}
\def\eKzKPi  {\ensuremath{\etac \to \Kz \Km \pip}\xspace}
\def\eKKbarPi {\ensuremath{\etac \to \KKbar \pi}\xspace}
\def\FourK  {\ensuremath{\KpKm \KpKm}\xspace}
\def\TwoPhi {\ensuremath{~\phi~\phi}\xspace}
\def\KKPi   {\ensuremath{\KpKm \piz}\xspace}
\def\KsKPi  {\ensuremath{\KS \Kpm \pimp}\xspace}

\def\KKbarPi {\ensuremath{\KKbar \pi}\xspace}
\long\def\inst#1{\par\nobreak\kern 4pt\nobreak
    {\it #1}\par\vskip 10pt plus 3pt minus 3pt}

\begin{document}
{\pagestyle{empty}

\begin{flushright}
\babar-\BaBarType-\BaBarYear/\BaBarNumber \\
SLAC-PUB-\SLACPubNumber \\
March, 2002
\end{flushright}

\par\vskip 3cm

% Title of the paper
\begin{center}
\Large \bf \boldmath
   Branching Fraction Measurements \\ of the Decays \etacK,
where \eKKbarPi\ and \eKKKK
\end{center}
\bigskip

\begin{center}
\large The \babar\ Collaboration\\
\mbox{ }\\
%\posted
%\draft
%\today
\end{center}
\bigskip \bigskip

% Abstract
\begin{center}
\large \bf Abstract
\end{center}

In this report, we present the observation of the exclusive
decays \etacKn\ and \etacKP, and the measurement of 
the related branching fractions. Using a sample of 22.7$\times 10^6$
\upsbb\ decays collected with the \babar\ detector at the SLAC \pep2\ 
\abf\ during 1999-2000, we have observed statistically significant
signals in the \eKsKPi\ and \KKPi\ channels and set upper limits in the
\eFourK\ channels. All the results presented are preliminary.

We have measured
\begin{tabbing}
xxxxxxxxxxxxxxxxxxxxxxx
\= \BR(\etacKP) \= = \= (1.59~\PM~0.20~\PM~0.13~\PM~0.49)$~10^{-3}$ \kill
\> \BR(\etacKP) \> = \> (1.50~\PM~0.19~\PM~0.15~\PM~0.46)$\times 10^{-3}$ \\
\> \BR(\etacKn) \> = \> (1.06~\PM~0.28~\PM~0.11~\PM~0.33)$\times 10^{-3}$
\end{tabbing}
where the first error is statistical, the second systematic and
the last due to the uncertainty on the world average \eKKbarPi\ branching 
fraction.
\vfill
Presented at the XXXVII$^{th}$ Rencontres de Moriond on
QCD and Hadronic Interactions,\\
\centerline{
3/16--3/23/2002, Les Arcs, Savoie, France
}
\newpage

}% End of pagestyle{empty}

% Input author list file
\begin{center}
\small

The \babar\ Collaboration,
\bigskip

%% author list as of 01-Mar-2002 (548 authors)
B.~Aubert,
D.~Boutigny,
J.-M.~Gaillard,
A.~Hicheur,
Y.~Karyotakis,
J.~P.~Lees,
P.~Robbe,
V.~Tisserand,
A.~Zghiche
\inst{Laboratoire de Physique des Particules, F-74941 Annecy-le-Vieux, France }
A.~Palano,
A.~Pompili
\inst{Universit\`a di Bari, Dipartimento di Fisica and INFN, I-70126 Bari, Italy }
% ...

\end{center}%\newpage

% reset footnote counter
\setcounter{footnote}{0}

% The body of the paper starts here
\section{Introduction}
\label{sec:Introduction}

We present the measurement of the branching fractions of the exclusive
decays\footnote{Throughout this paper,
whenever a mode is given, the charge conjugate (c.c.) is also implied.}
\etacKn and \etacKP,
with \etac\ decaying into \KsKPi, \KKPi, and \FourK (\ksdk\ and \pizdk).
The \etac\ is a \ccbar\ meson with $I^G(J^{PC}) = 0^+(0^{-+})$.
The decay \etacKn\ proceeds through the same \b\ \to\ \ccbar \s\ 
color-suppressed quark diagram as the ``golden''
mode, \golden, used to measure the  \CP--violating parameter
\stwob\ with negligible theoretical uncertainty~\cite{ref:sin2b}.
Up to now, experimental information on 
\B\ decays into \etac\ has been sparse~\cite{ref:CLEOincl, ref:CLEOexcl}.

The ratio of the decay rates for the exclusive charmonium decays
\begin{equation}
  R_K\;\equiv\;\Gamma(B \rightarrow \etac K)/\Gamma(B \rightarrow \jpsi K)
\end{equation}
has been calculated with different dynamical 
assumptions~\cite{ref:Ahmady1}--\cite{ref:Hwang} including
factorization\footnote{We note that we
have measured a departure from the factorization hypothesis~\cite{ref:babarjpsiKstar}
in another \b\ \to\ \ccbar \s\ color-suppressed mode, \jpsiKstar, 
wherein we have made a
polarization measurement, more sensitive than a measurement of $R_K$
to the existence of a factorization-violating term.}.
The ratio is used since one expects that the corrections to the heavy
quark limit, due to the relatively light \s-quark, are likely to cancel.
This leads to the following predictions for $R_K$:
1.6~\PM~0.2~\cite{ref:Ahmady1},
1.64~\PM~0.55~\cite{ref:Deshpande},
1.8~$\sim$~2.3~\cite{ref:Gourdin},
0.94~\PM~0.25~\cite{ref:Colangelo},
1.0~$\sim$~1.3~\cite{ref:Hwang}.

\section{\boldmath The \babar\ detector and dataset}
\label{sec:babar}

The data used in this analysis are obtained with the \babar\ detector
at the \pep2\ asymmetric \epem\ storage ring. 
The \babar\ detector is described elsewhere~\cite{ref:babarnim}.
The 1.5~T superconducting solenoidal magnet, whose cylindrical volume is 
$\approx 1.4$\m in radius and $\approx 3$\m\ long, contains a
charged-particle tracking
system, a Cherenkov detector (DIRC) dedicated to charged particle identification 
and an electromagnetic
calorimeter. The segmented iron flux return, including endcaps, 
provides identification of muons
and \KL. In addition, the end of the cylindrical volume in the \en\ direction 
is instrumented with an
electromagnetic calorimeter. The tracking system consists of a 5-layer double-sided
silicon vertex tracker and a 40-layer drift chamber filled with a gas mixture
of helium and isobutane. The calorimeter consists of 6580 CsI(Tl) crystals. The
flux return is instrumented with resistive plate chambers.

We have used data corresponding to 20.7 \invfb\ of integrated luminosity 
collected at the \FourS\ resonance (``on-resonance"), and
2.1~\invfb\ recorded (``off-resonance") about $40$\mev\ lower in energy in the \FourS\ rest
frame (``CM"), between October 1999 and October 2000. The asymmetric collisions
produce a boost in the \en\ direction, with $\beta \gamma = 0.55$ in on-resonance
running. 

\section{\boldmath Analysis method}
\label{sec:Analysis}

A blind analysis is performed in which all selections are chosen to
maximize $N_S/\sqrt{N_S+N_B}$ using
simulated or off-resonance data, or sidebands in on-resonance data.
$N_S (N_B)$ is the number of expected signal (background) events after all selection criteria have
been applied.

Event selection designed
to enhance the number of \B\ decays requires
four or more charged tracks,
the sum of all charged and neutral energies to be above 2~\gev,
the sum of all the charged momenta to be above 1~\gevc, and
the normalized second Fox-Wolfram moment~\cite{ref:R2} to be less than 0.6.
In addition, at least one neutral or charged kaon candidate is required to have a momentum
in the \FourS\ rest frame consistent with the two-body decay \etacK.

\Bz\ or \Bpm candidates are formed from an \etac\ candidate and a ``fast" kaon,
either a charged kaon or a \ksdk.
The \etac\ candidates correspond to three different topologies: two charged
tracks with either \ksdk\ or \pizdk, or four charged tracks. The \B\ decay vertex is
calculated using the charged \etac\ daughters, and the fast kaon if charged.

We require that all charged tracks be within
$0.35 < \theta < 2.54$ to obtain well-reconstructed tracks, where $\theta$ is the polar
angle with respect to the \en\ direction.
An important requirement of our analysis is that charged kaon
candidates from the \etacK\ decay are identified by the
DIRC and/or by measurements of
ionization energy loss $dE/dx$ in the drift chamber and silicon tracker.
The momentum of each kaon from \etac\ decay is required to be greater than 250\mevc.

The \KS\ particles can arise from \etac\ or $B$ decays.
In the following, the number in parentheses corresponds to the latter \KS.
The \ksdk\ candidates are required to have a reconstructed invariant mass within 
12.5~(10)~\mevcc\ of the nominal, i.e. world average, value~\cite{ref:pdg2000}.
Furthermore, the cosine of the opening
angle between the flight direction and the momentum vector of the \KS\ candidate
is required to be greater than 0.990~(0.9995), and the flight distance from the \B\
vertex greater than 2~(3) times its error.

The \pizdk\ candidates are formed from pairs of photons detected in the calorimeter
with a reconstructed invariant mass within 15\mevcc\ of the nominal value. We 
require that the
cosine of the decay angle in the \piz\ rest frame be less than 0.82
to avoid accidental combinations involving very soft photons.
In addition, the electromagnetic showers are required to have moments of the 
lateral energy deposition~\cite{ref:LAT} 
between 0.01 and 0.55. The lower energy photon has a minimum
energy of 130\mev\ while the minimum value for the higher energy photon is
270\mev.

A Fisher discriminant is used to suppress continuum backgrounds. 
The Fisher variable is defined as a linear combination of eighteen variables,
including the energies between each of nine cones relative to
the \etac\ direction in the CM~\cite{ref:cones}.
The most important variables are the normalized second Fox-Wolfram moment 
and the event thrust, constructed with all charged tracks and neutral clusters in the event.
The Fisher discriminant is trained on signal, \uubar, \ddbar, \ssbar,
and \ccbar\ simulated events, and tested on off-resonance data.
The requirements on the Fisher discriminant depends on the decay mode.

The charmonium mass region is defined by 2.74~$<m_X<$~3.22~\gevcc. After all
selection criteria, the weighted double-Gaussian 
mass resolutions are 10, 12, and 26~\mevcc\ for the \KsKPi, \FourK\ and \KKPi\ 
channels, respectively, as obtained from a simulation.
The \etac\ signal region
varies between $\pm55$\mevcc\ and $\pm70$\mevcc\ relative to the nominal \etac\ mass
(2979.8\mevcc), depending on the \etac\ decay mode.

The total energy of the \epem\ system in the \FourS\ CM 
and laboratory frames are denoted by $\sqrt{s}$ and $E_o$, respectively.
In the \epem\ laboratory frame, the candidate energy is defined as
$E_B=(s/2+{\bf p}_o\cdot{\bf p}_B)/E_o$,
where ${\bf p}_o$ and ${\bf p}_B$ are
the momentum vectors of the \epem\ system and the
\B\ candidate, respectively~\cite{ref:babarnim}.
The analysis region is defined by a rectangular area 
in the \DeltaE--\mes\ plane where \DeltaE\
is the difference between the energy of the \B\ candidate  
in the CM frame and $\sqrt{s}/2$, 
and \mes\ is the beam-energy substituted mass, $\sqrt{E_B^2 - p_B^2}$.
For events with multiple candidates, the one with the smallest
\absDe\ is retained; this choice affects only a small fraction of events,
from 3.4\% to 12.4\% in the analysis region.

The limits of the analysis region are defined by 5.1$<$\mes$<$5.29~\gevcc 
and \absDe$<$0.25~\gev. 
According to the full detector simulation based on GEANT3~\cite{ref:GEANT}, 
depending on the \etac\ decay mode. 
the signal is Gaussian-distributed in \DeltaE\ with
a mean near zero and a resolution between 15 and 30\mev, while 
it is Gaussian-distributed in \mes\ with
a mean near the $B$ mass and a resolution around 2.5\mevcc.
The \DeltaE\ resolution depends on the
\etac\ decay mode, best for \KsKPi\ and worst for \KKPi. 
Note that the \DeltaE\
distribution in data is not centered at zero but rather at about $-10$\mev;
the window is shifted accordingly, leading to a contribution to the overall
systematic error. The shifted-\absDe,\absmes\ signal region is
$<$30~\mev, $<$7~\mevcc for the tightest (\KsKPi) case and
$<$70~\mev, $<$9~\mevcc for the loosest (\KKPi) case.

\section{\boldmath Observation of exclusive \etac\ signals}
\label{sec:Observation}
\begin{figure}[!htb]
\begin{center}
\includegraphics[height=8cm]{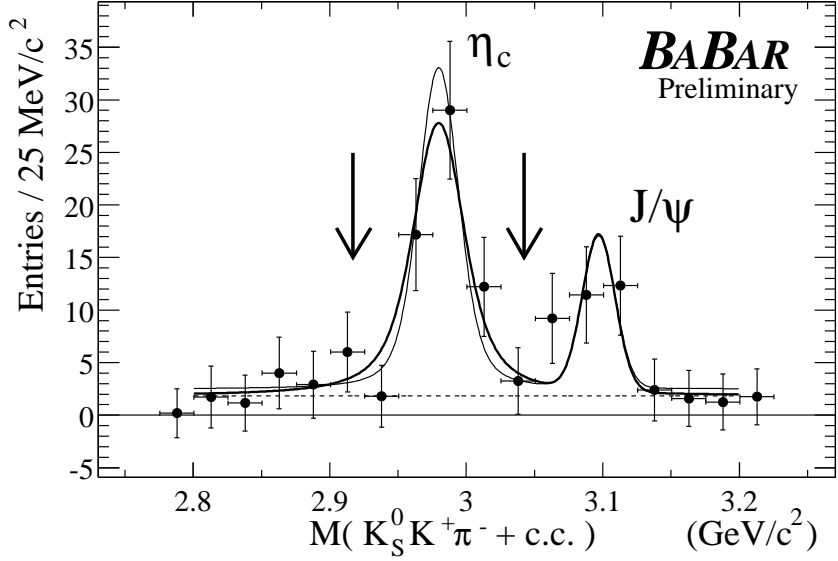}
\caption{\footnotesize{\KsKPi\ mass
for \Bp\ candidates in the (\DeltaE, \mes)~signal region after subtraction 
of the combinatorial background. 
The fits are described in the text.
The remaining flat background is that 
due to the peaking background; see text.
The ``thick" curve corresponds to
the CLEO \etac\ width while 
the ``thin" curve corresponds to
the world average.
}}
\label{fig:etacmass}
\end{center}
\end{figure}

Figure~\ref{fig:etacmass} displays the mass distribution of the
charmonium system in the (\DeltaE,\mes)~signal region for the
\KsKPi\ channel, using \Bp\ candidates after subtraction of the 
combinatorial background. We see clear \etac\ and \jpsi\ peaks where
we have indicated the \etac\ mass selection excluding the \jpsi\ region.
The representative curves are fits of three contributions: 
flat background, \jpsi\ peak, and \etac\ peak with two different
widths. The \jpsi\ peak is represented by a 
Gaussian with mean constrained at the nominal \jpsi\ mass and a 
12~\mevcc\ resolution. The \etac\ mass peak is represented by a Breit-Wigner
distribution convoluted with the same  Gaussian. The
mean of the Breit-Wigner distribution is fixed at the nominal
\etac\ mass and the width\footnote{
The world average width is
$13.2^{+3.8}_{-3.2}$\mevcc\ while more recent results give
$27.0 \pm 5.8 \pm 1.4$\mevcc~\cite{ref:CLEOwidth},
$11.0 \pm 8.1 \pm 4.1$\mevcc~\cite{ref:BESwidth} and
$21.1^{+6.9}_{-6.2}$\mevcc~\cite{ref:E835width}},
either
to the world average or to the CLEO measurement~\cite{ref:CLEOwidth}.
Since we cannot yet distinguish among the various measurements,
we have used for the \etac\ width the average value $16.7 \pm 6.0$\mevcc;
the efficiency depends on the width central value and the systematic error
on its error.

\begin{figure}[!htb]
\begin{center}
\begin{tabular}{cc}
\includegraphics[height=8cm]{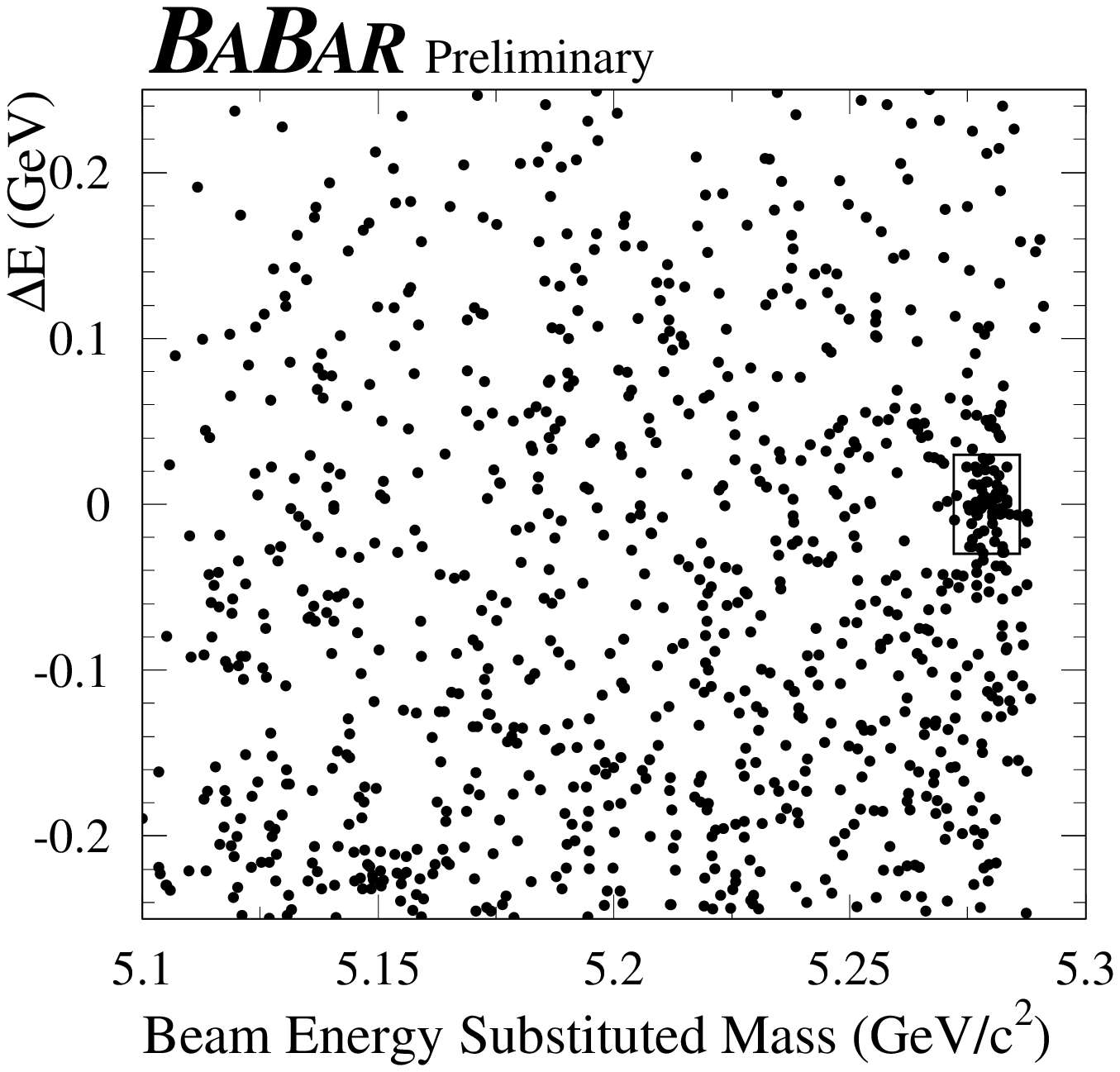} &
\includegraphics[height=8cm]{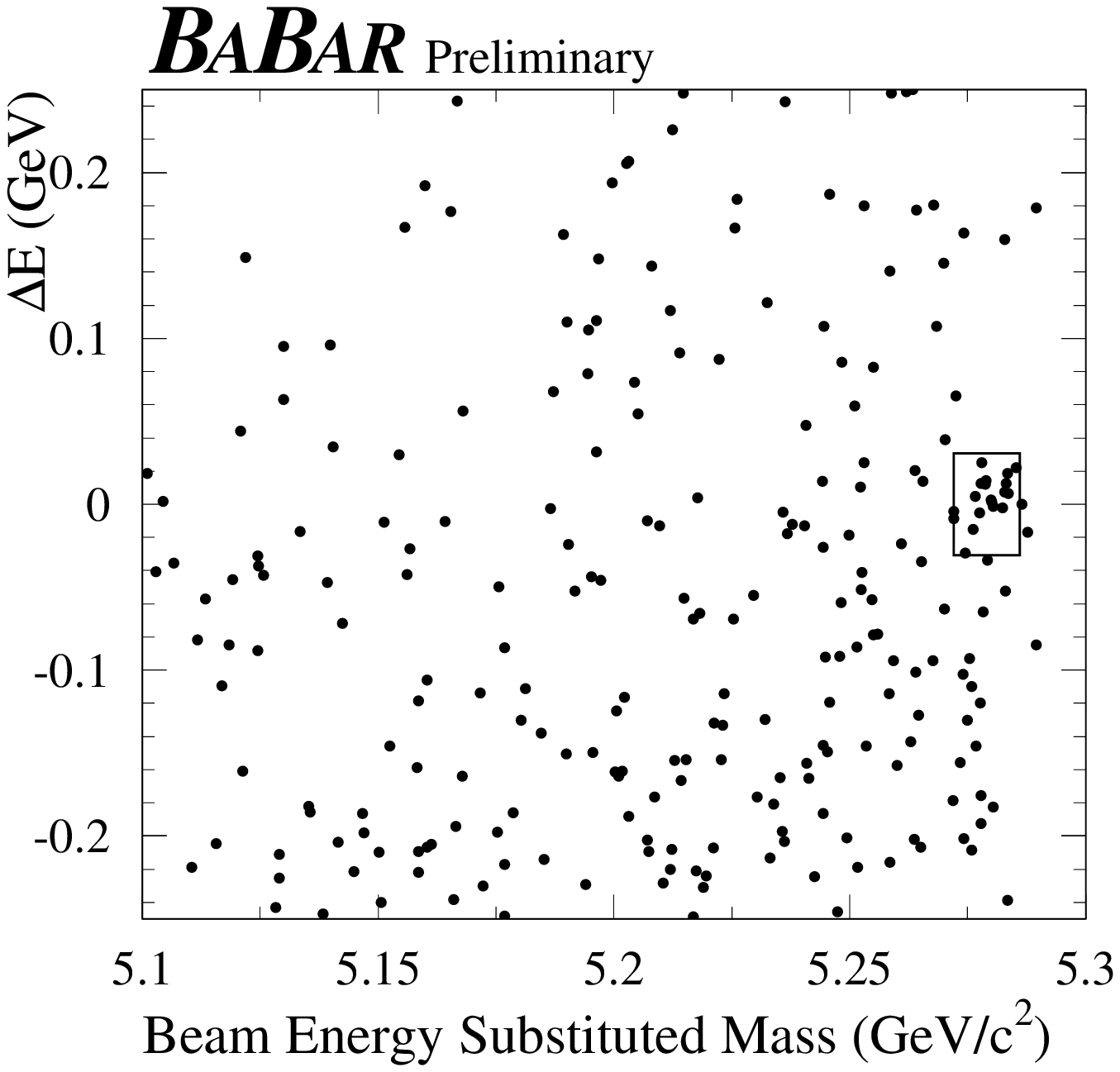}
\end{tabular}
\caption{\footnotesize{\DeltaE\ {\it vs.} \mes\ for 
candidate \etacKP\ events (on left)
and \etacKN\ events (on right), with \eKsKPi. The (\DeltaE,\mes)~signal region is indicated.
All selection criteria have been applied except for the signal region requirements.
}}
\label{fig:analysisbox}
\end{center}
\end{figure}

\begin{figure}[!htb]
\begin{center}
\includegraphics[height=8cm]{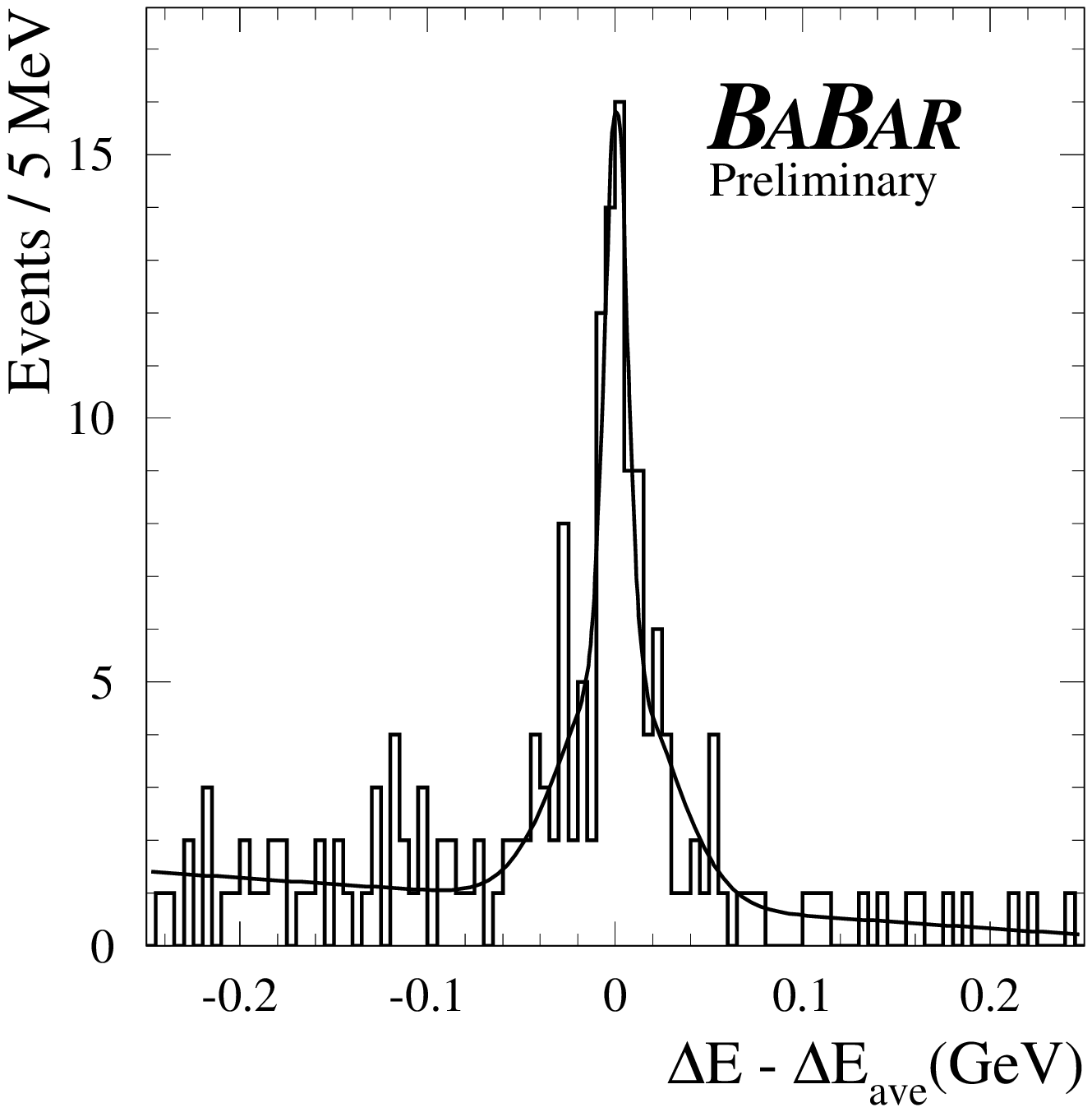}
\caption{\footnotesize{The
\DeltaE\ distribution relative to its mean in the \mes\ signal band for
combined \etacKP\ and \etacKN\ candidates with \eKsKPi, fitted to
a double Gaussian with common mean on top of a linear background. The 
weighted average resolution is 16.3~\mevcc.
The narrower Gaussian represents 71\% of the area of the
double Gaussian; its resolution is 5.9~\mevcc.
All selection criteria have been applied except that for \DeltaE.}}
\label{fig:deltaE}
\end{center}
\end{figure}

\begin{figure}[!htb]
\begin{center}
\begin{tabular}{cc}
\includegraphics[height=8cm]{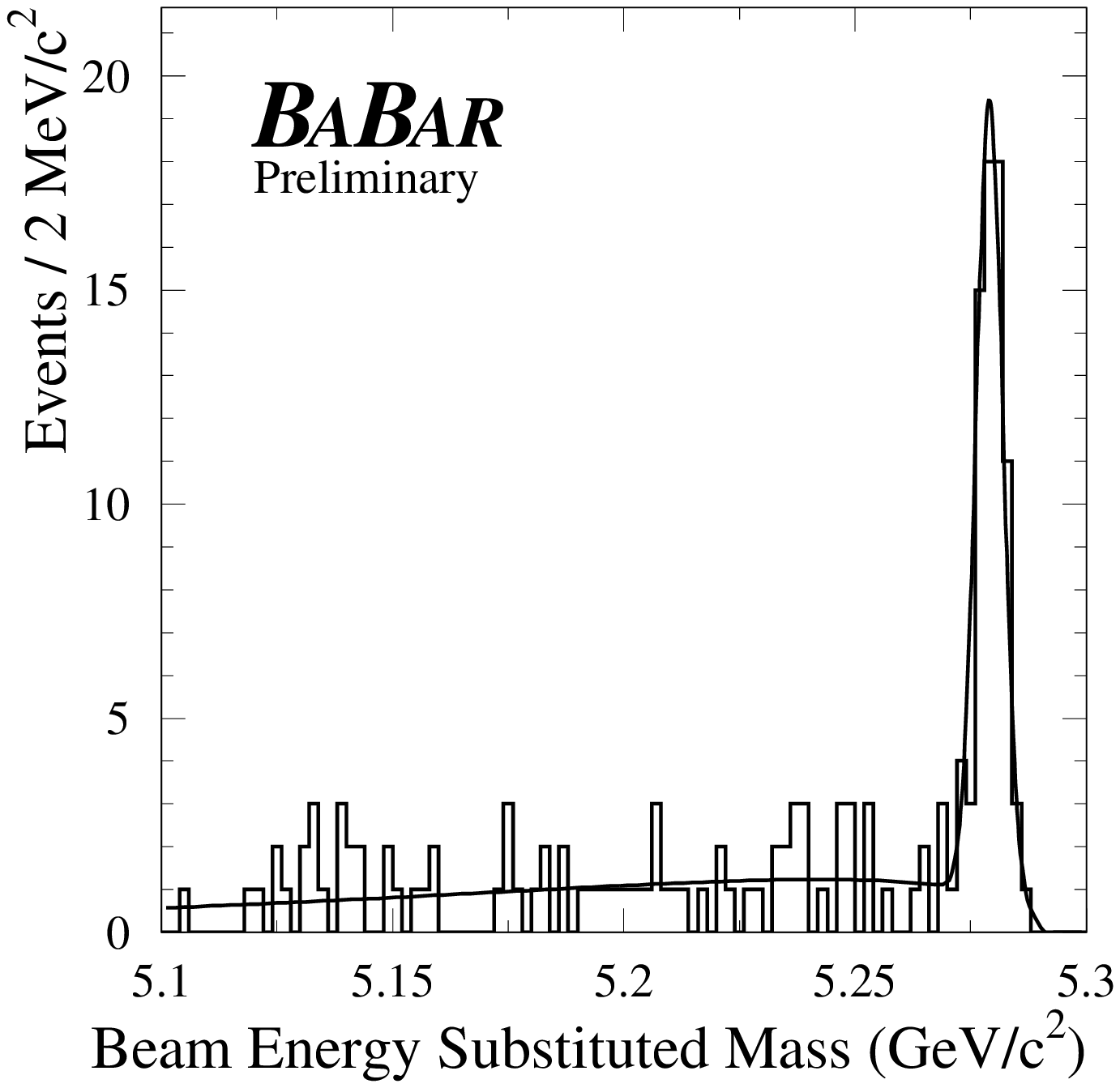}
                  &
\includegraphics[height=8cm]{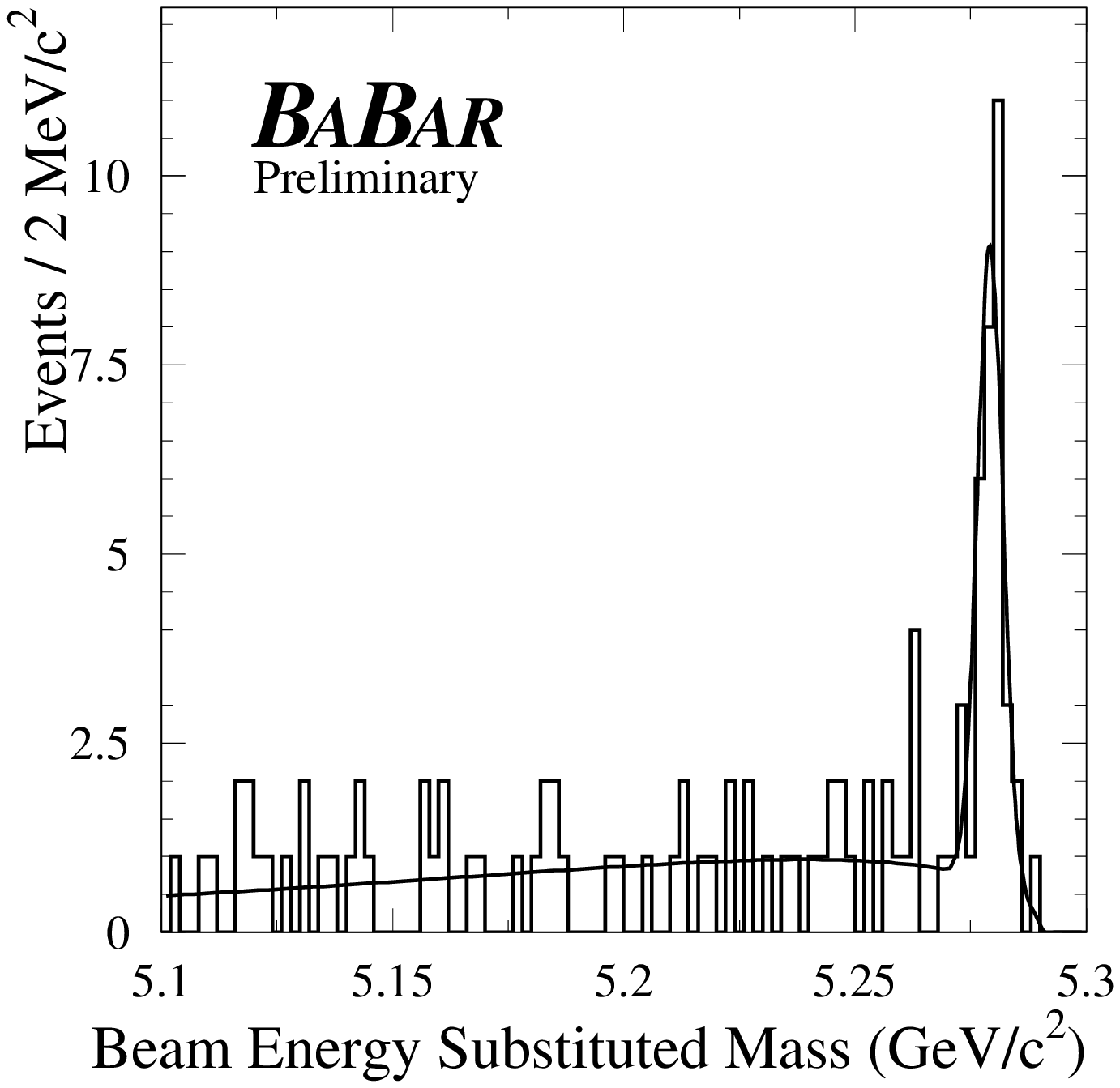}
\end{tabular}
\begin{tabular}{c}
\includegraphics[height=8cm]{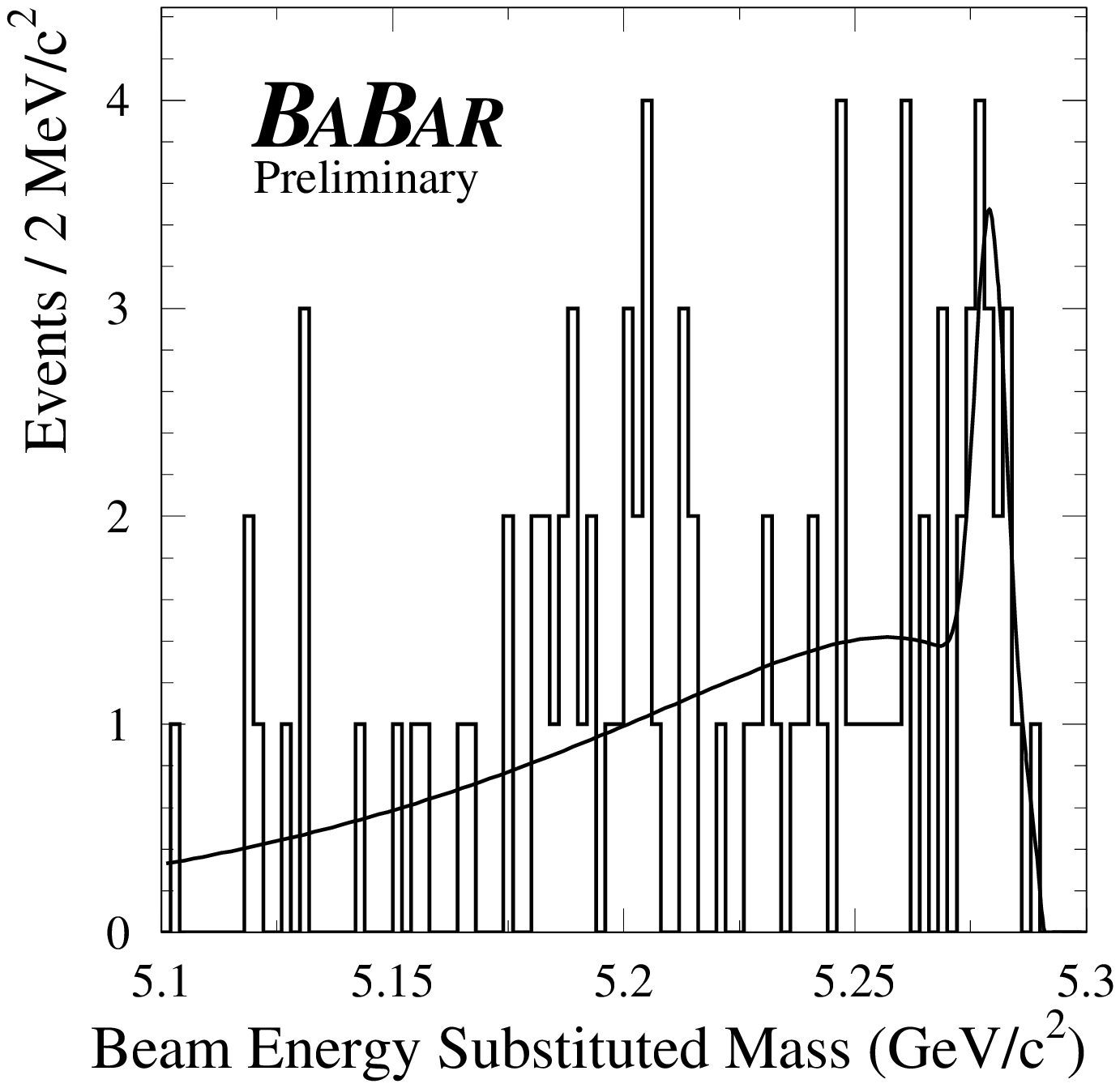}
\end{tabular}
\caption{\footnotesize{\mes\ distributions in the \DeltaE\ signal band for candidate \etac\ decays:
\KsKPi\ (top left), \KKPi\ (top right), and \FourK\ (bottom).
All plots use combined \Bp\ and \Bz\ data and
are fitted, using a binned likelihood method, to the sum of
a Gaussian signal, with an average resolution of 2.9~\mevcc,
and a threshold background function~\cite{ref:argus}.
All selection criteria have been applied except that \mes.}}
\label{fig:mes}
\end{center}
\end{figure}

In Figure~\ref{fig:analysisbox} we display the analysis region for
the \Bp\ and \Bz(\eKsKPi) channels as examples. Clear accumulations in the 
(\DeltaE,\mes)~signal region are apparent.
Figures~\ref{fig:deltaE} and \ref{fig:mes} display projections of the analysis region for
the different \etac\ channels.
The combinatorial background shape is parametrized by a linear function in 
\DeltaE\ and a threshold function~\cite{ref:argus} in \mes\ with a fixed endpoint
given by the average beam energy. 

In addition to the combinatorial background, a background 
that peaks in the (\DeltaE,\mes)~signal region can arise
from cross-feed from other \etac\ decay modes, from partial reconstruction
and/or incorrect particle identification, or from \B\
decays into the same detected particles without an intermediate \etac\ decay
(exact matches).
After study, the first two sources are found to be negligible. A quantitative
evaluation of the exact matches for each mode is made using data by studying
the \etac\ mass sidebands for events in the (\DeltaE,\mes)~signal region, after subtracting
the combinatorial background as a function of mass.
The peaking
background is consistent with zero for all modes except possibly for
the \KsKPi\ mode, see the flat background in Figure~\ref{fig:etacmass}.

The raw yield and expected backgrounds in the (\DeltaE,\mes)~signal region, and the probability that the
background fluctuates to the observed yield are given in Table~\ref{tab:yield}.
In order to ensure the statistical independence of the
signal and background measurements, the combinatorial background is estimated here by
the extrapolation into the \mes\ signal band of the threshold function
fitted in the \DeltaE\ signal band below the \mes\ signal band (\mes~$<
5.27$\gevcc). Because of the low statistics in the \Bz\
channels, the shape parameter of the background function is fixed to
that fitted in the corresponding \Bp\ channel.
%----------------------------------begin table
\begin{table}[!htb]
\caption{\footnotesize{Raw yield, extrapolated combinatorial (see text)
and peaking backgrounds in the (\DeltaE,\mes)~signal region,
and Poisson probability that the combined background
fluctuates to the number of events found in the signal region (called 
``Prob$_{fluct}$"). Due to the limited data sample,
the fitted combinatorial background estimate for \Bz(\eFourK)
comes from the \DeltaE\ sidebands.}}
\footnotesize{
\label{tab:yield}
\bigskip\noindent
\begin{center}
\begin{tabular}{c|c|c|c|c}\hline
mode           &Yield    & Fitted      & Peaking      &Prob$_{fluct}$  \\
               &         &combinatorial&background    &                \\
               &         &background   &              &                \\ \hline \hline
\Bp(\eKsKPi)   &  72     & 6.08 \PM\ 1.39 & 6.12 \PM\ 2.61  & $2 \times 10^{-16}$\\ \hline
\Bp(\eKKPi)    &  25     & 2.92 \PM\ 0.92 & 0.58 \PM\ 0.58  & $3 \times 10^{-15}$\\ \hline
\Bp(\eFourK)   &  17     & 7.41 \PM\ 1.78 & 1.72 \PM\ 2.75  & $2 \times 10^{-3}$ \\ \hline \hline
\Bz(\eKsKPi)   &  19     & 1.18 \PM\ 0.38 & 1.48 \PM\ 1.08  & $3 \times 10^{-13}$ \\ \hline
\Bz(\eKKPi)    &   8     & 1.73 \PM\ 0.38 & 0               & $4 \times 10^{-4}$ \\ \hline
\Bz(\eFourK)   &   1     & 1.01 \PM\ 0.25 & -               & -              \\ \hline
\end{tabular}
\end{center}
}
\end{table}
%------------------------------------end table

\section{\boldmath Branching fraction determination}
\label{sec:BF}
\par
The measured \Bp\ or \Bz\ branching fraction (\BR) is given by
\begin{equation}
  \BR = { N_{yield} \over {N_{\BB} \times \epsilon} } ,
\end{equation}
where $N_{yield}$ is the net yield in the (\DeltaE,\mes)~signal region,
extracted from fits to the \mes\ distributions in the \DeltaE\ signal
region (Figure~\ref{fig:mes}),  and corrected for the peaking background contributions
listed in Table~\ref{tab:yield};
$\epsilon$ is the signal efficiency determined
by applying the same analysis chain to signal
Monte Carlo (MC) samples and correcting for data-MC differences; and $N_{\BB}$ is 
the number of produced \BB\ pairs, (22.73\PM~0.36)$\times 10^6$, determined
by a comparison of the rate of multihadron events taken on-resonance to that
off-resonance.

\subsection{\boldmath Determination of signal efficiency}
\label{sec:Efficiency}
The efficiency for reconstructing \etacK\ candidates for each \etac\ decay
mode is given by the fraction of generated signal events that are 
reconstructed
in the appropriate mode. We have compared simulations with real data, using
for example \tautau and \Dstarpm\ control samples.
There are small differences in reconstruction efficiency for 
charged particles, \KS\ and \piz mesons,
vertexing efficiency, resolution and absolute scale of charged particle momentum
and photon energies, and charged kaon identification and pion misidentification
probabilities. These effects have been measured and corrected.
The resulting efficiencies are given in the first line of Table~\ref{tab:effsyst}.
%----------------------------------begin table
\begin{table}[!htb]
\caption{\footnotesize{Relative systematic errors on efficiency.
All values are expressed in percentage relative to the efficiency,
which is given in the first line as a fraction. The last line
gives the total relative systematic error obtained as a sum in
quadrature of the individual contributions. The 1.6\% error from 
the determination of the number of \BB\ events, common to all modes, 
is not listed but is included in the total as is the statistical
error on the efficiency determination.}}
\footnotesize{
\label{tab:effsyst}
\bigskip\noindent
\begin{center}
\begin{tabular}{c|r|r|r|r|r|r}\hline
                     & \etacKN    & \etacKN   & \etacKN & \etacKP    & \etacKP   & \etacKP\\
\etac\ decay         & \FourK     & \KsKPi    & \KKPi   & \FourK     & \KsKPi    & \KKPi  \\ \hline \hline
Efficiency           & 0.111      & 0.148     & 0.0733  & 0.117      & 0.145     & 0.0635 \\ \hline
Rel. stat. err.      & 0.004      & 0.003     & 0.0027  & 0.003      & 0.003     & 0.0017 \\ \hline \hline
Tracking eff.        & 9.5        & 7.7       & 5.8     & 7.8        & 6.2       & 4.4    \\ \hline
\KS\ eff. and cuts   & 5.9        & 12.2      & 5.3     &  -         & 6.8       &  -     \\ \hline
\g eff. and \piz\ cuts & -        &  -        & 3.5     &  -         &  -        & 3.5    \\ \hline
Vertexing eff.       & 1.3        & 1.0       & 1.0     & 1.3        & 1.0       & 1.0    \\ \hline
Kaon ident. eff.     & 10.5       & 2.9       & 5.6     & 14.1       & 6.5       & 9.2    \\ \hline
Fisher cut eff.      & 2.3        & 2.2       & 4.0     & 1.1        & 2.9       & 4.0    \\ \hline
\etac\ width uncert. & 2.9        & 3.2       & 3.1     & 2.9        & 3.2       & 3.1    \\ \hline
\DeltaE\ centroid shift  & 0.47   & 3.3       & 0.86    & 0.51       & 2.7       & 0.24   \\ \hline
\DeltaE\ resolution  & 3.6        & 3.8       & 4.4     & 3.4        & 5.2       & 3.4    \\ \hline \hline
$\Sigma$             & 16.2       & 16.1      & 12.3    & 16.8       & 13.5      & 12.4   \\ \hline \hline
\end{tabular}
\end{center}
}
\end{table}
%------------------------------------end table

\subsection{\boldmath Determination of systematic errors}
\label{sec:Systematics}
We have evaluated the systematic errors on the yield, \B~counting and
efficiency determination. The systematic error on the yield comes from
a comparison of the combinatorial background estimations from the \DeltaE\
side and signal bands while that on \B~counting comes principally
from the uncertainty on the efficiency due to small differences between
data and simulation.

Each of the efficiency corrections, as well as our knowledge of the \etac\ width, 
has a corresponding systematic uncertainty. In addition,
each requirement in the analysis method has been studied to evaluate any systematic
differences between simulation and data.
The dominant systematic errors on the signal efficiency are due to kaon identification,
tracking efficiency, and \KS\ reconstruction as can be seen in 
Table~\ref{tab:effsyst}.

\subsection{\boldmath Results}
\label{sec:Results}
Our results for the product
of the branching fractions for each mode are listed below.
We have used the nominal values for the \ksdk\ and \pizdk\
branching fractions.
The \etacK\ branching-fraction
determinations assume that the branching
fraction of the \FourS\ into \BB\ is 100\%, with an equal admixture of charged 
and  neutral $B$ final states, and similarly for \Kz\ relative to \KS\ and \KL.
\begin{tabbing}xxxxxxxxx
\= \BR(\etacKP)\BR(\eKzKPi\ + c.c.) \=  =  \= (52.8\PM 7.9\PM 7.3)$\times 10^{-6}$\kill
\> \BR(\etacKP)\BR(\eKzKPi\ + c.c.) \>  =  \> (52.8\PM 7.9\PM 7.3)$\times 10^{-6}$\\
\> \BR(\etacKP)\BR(\eKKPi) \> = \> (15.5\PM 3.6\PM 2.5)$\times 10^{-6}$\\
\> \BR(\etacKP)\BR(\eFourK) \> $<$ \> 5.6$\times 10^{-6}$ (90\% CL)\\
\> \BR(\etacKn)\BR(\eKzKPi\ + c.c.) \> = \> (36.8\PM 11.6\PM 6.0)$\times 10^{-6}$\\
\> \BR(\etacKn)\BR(\eKKPi)  \> = \> (11.3\PM 5.1\PM 2.4)$\times 10^{-6}$\\
\> \BR(\etacKn)\BR(\eFourK) \> $<$ \> 2.3$\times 10^{-6}$ (90\% CL)
\end{tabbing}
The first error is statistical and the second systematic.
The central value for \BR(\etacKP)\BR(\eFourK) is 3.2~$\times 10^{-6}$,
while the two-sided 68\% CL varies from 2.6$\times 10^{-6}$ to 4.1$\times 10^{-6}$.
No correction is made for any potential \TwoPhi\ contribution to the \FourK\
channels.

\par
The channels \eKsKPi\ and \eKKPi\ are manifestations of the
general decay \eKKbarPi. From isospin symmetry,
the corresponding rates are related by simple Clebsch-Gordon 
coefficients: \BR(\eKzKPi+ c.c.)~=~2/3~\BR(\eKKbarPi) and \BR(\eKKPi)~=~1/6~\BR(\eKKbarPi).
Therefore the ratio of branching fractions, \BR(\eKKPi)/\BR(\eKzKPi+ c.c.), should be
0.25. Our measurements are consistent with this value for \Bp\ (0.29~\PM~0.08~\PM~0.04)
and \Bz\ (0.31~\PM~0.17~\PM~0.05).

We therefore combine our two results, taking into account common systematic errors,
to obtain the values for the general decay:
\begin{tabbing}
xxxxxxxxxxxxxxxxx
\= \BR(\etacKP)\BR(\eKKbarPi) \= = \= (87.5~\PM~11.1~\PM~7.3)$\times 10^{-6}.$\kill
\> \BR(\etacKP)\BR(\eKKbarPi) \> = \> (82.5~\PM~10.4~\PM~8.3)$\times 10^{-6}$\\
\> \BR(\etacKn)\BR(\eKKbarPi) \> = \> (58.1~\PM~15.2~\PM~6.3)$\times 10^{-6}.$
\end{tabbing}
The first error is statistical and the second systematic.
We deduce the branching fraction ratio from our measurements of the \KKbarPi\ channel:
\BR(\etacKn)/\BR(\etacKP)~=~0.71~\PM~0.20~\PM~0.08.
We have not used the \eFourK\ results since their statistical weight would be marginal.

\par
Using the world average for the \eKKbarPi\ branching fraction,
0.055 \PM\ 0.017~\cite{ref:pdg2000}, our results become
\begin{tabbing}
xxxxxxxxxxxxxxxxxxxxxx
\= \BR(\etacKP) \= = \= (1.59~\PM~0.20~\PM~0.13~\PM~0.49)$\times 10^{-3},$ \kill
\> \BR(\etacKP) \> = \> (1.50~\PM~0.19~\PM~0.15~\PM~0.46)$\times 10^{-3}$ \\
\> \BR(\etacKn) \> = \> (1.06~\PM~0.28~\PM~0.11~\PM~0.33)$\times 10^{-3},$
\end{tabbing}
where the last error is due to the \eKKbarPi\ branching fraction.
We have not used the \eFourK\ results since the \etac\ branching fraction is
not very well known.
We compare these results to the
exclusive branching fractions measured by CLEO~\cite{ref:CLEOexcl}:
\BR(\etacKP)~=~($0.69^{+0.26}_{-0.21} \pm 0.08 \pm 0.20)\times 10^{-3}$
and
\BR(\etacKn)~=~($1.09^{+0.55}_{-0.42} \pm 0.12 \pm 0.31)\times 10^{-3}$
The third error is that due to the nominal \eGamEtac\ branching fraction.
Assuming that the errors due to the
nominal branching fractions cancel, our results
differ by a factor 2.2~\PM~0.9 for the \Bp\ channel,
combining statistical and systematic errors in quadrature. 
The \Bz\ channel results are consistent.

To determine $R_K$, we have used our measurements~\cite{ref:babarjpsiK}
of the branching fractions,  
\BR(\goldKP) = ($10.1 \pm 0.3 \pm 0.5)\times 10^{-4}$ and
\BR(\golden) =  ($8.5 \pm 0.5 \pm 0.6)\times 10^{-4}$, 
taking into account common systematic errors, to obtain
\begin{tabbing}
xxxxxxxxx
\= $R_K^+ =$ \= $\Gamma$(\etacKP)/$\Gamma$(\goldKP) \= = \= 
1.48~\PM~0.19~\PM~0.17~\PM~0.46 \kill
\> $R_K^+ =$ \= $\Gamma$(\etacKP)/$\Gamma$(\goldKP) \> = \> 
1.48~\PM~0.19~\PM~0.17~\PM~0.46 \\
\> $R_K^0 =$ \= $\Gamma$(\etacKn)/$\Gamma$(\golden) \> = \> 
1.24~\PM~0.33~\PM~0.16~\PM~0.38 \\
\end{tabbing}
where the first error is statistical, the second systematic and the third
due to the \eKKbarPi\ branching fraction. Our results agree with the 
theoretical predictions listed at the end of Section 1.

\section{Acknowledgments}
\label{sec:Acknowledgments}

% Standard acknowledgments paragraph; must always be included.
We are grateful for the 
extraordinary contributions of our \pep2\ colleagues in
achieving the excellent luminosity and machine conditions
that have made this work possible.
The success of this project also relies critically on the 
expertise and dedication of the computing organizations that 
support \babar.
The collaborating institutions wish to thank 
SLAC for its support and the kind hospitality extended to them. 
This work is supported by the
US Department of Energy
and National Science Foundation, the
Natural Sciences and Engineering Research Council (Canada),
Institute of High Energy Physics (China), the
Commissariat \`a l'Energie Atomique and
Institut National de Physique Nucl\'eaire et de Physique des Particules
(France), the
Bundesministerium f\"ur Bildung und Forschung
(Germany), the
Istituto Nazionale di Fisica Nucleare (Italy),
the Research Council of Norway, the
Ministry of Science and Technology of the Russian Federation, and the
Particle Physics and Astronomy Research Council (United Kingdom). 
Individuals have received support from 
the A. P. Sloan Foundation, 
the Research Corporation,
and the Alexander von Humboldt Foundation.

\end{document}